\documentclass[10pt,final,conference,letterpaper]{IEEEtran}

\usepackage{cite}

\ifCLASSINFOpdf{}
  \usepackage[pdftex]{graphicx}
   \graphicspath{{./pdf/}{./jpeg/}{./png/}}
   \DeclareGraphicsExtensions{.pdf,.jpeg,.png}
\else
\fi
\usepackage[caption=false,font=footnotesize]{subfig}
\usepackage{textgreek}
\usepackage{textcomp}

\usepackage{lineno,hyperref}

\usepackage{amsmath}
\begin{document}
%
\title{{\large \textcopyright 2018 IEEE. Personal use of this material is permitted. Permission from IEEE must be obtained for all other uses, in any current or future media, including reprinting/republishing this material for advertising or promotional purposes, creating new collective works, for resale or redistribution to servers or lists, or reuse of any copyrighted component of this work in other works. DOI: 10.1109/CCECE.2018.8447626\\ .\\ O. Raad, M. Makdessi, Y. Mohamad, and I. Damaj, SysMART Indoor Services: A System of Connected and Smart Supermarkets, The $31^{st}$ Canadian Conference on Electrical and Computer Engineering, IEEE, Quebec City, Quebec, Canada, May 13\textendash16, 2018. P. 1\textendash6.\\  \url{https://doi.org/10.1109/CCECE.2018.8447626}}\\ SysMART Indoor Services: A System of Smart and Connected Supermarkets}

\author{\IEEEauthorblockN{Omar Raad, Majd Makdessi, Yazan Mohamad, and Issam Damaj}
\IEEEauthorblockA{Electrical and Computer Engineering Department\\
American University of Kuwait\\
Salmiya, Kuwait\\
Email: \{S00026922,S00027923,S00026193,idamaj\}@auk.edu.kw}}


\maketitle

\begin{abstract}
Smart gadgets are being embedded almost in every aspect of our lives. From smart cities to smart watches, modern industries are increasingly supporting the Internet-of-Things (IoT). SysMART aims at making supermarkets smart, productive, and with a touch of modern lifestyle. While similar implementations to improve the shopping experience exists, they tend mainly to replace the shopping activity at the store with online shopping. Although online shopping reduces time and effort, it deprives customers from enjoying the experience. SysMART relies on cutting-edge devices and technology to simplify and reduce the time required during grocery shopping inside the supermarket. In addition, the system monitors and maintains perishable products in good condition suitable for human consumption. SysMART is built using state-of-the-art technologies that support rapid prototyping and precision data acquisition. The selected development environment is LabVIEW with its world-class interfacing libraries. The paper comprises a detailed system description, development strategy, interface design, software engineering, and a thorough analysis and evaluation. 
\end{abstract}


%
\IEEEpeerreviewmaketitle{}

\section{Introduction}
Shopping is one of the most frequent activities in today's busy schedule. Finding the required items inside a supermarket can prove to be a hurdle, especially, in hyper- and super-markets. Although going from one section to another can provide a healthy amount of walk for the day, arriving to the item's location and discovering that it ran out of stock is sometimes annoying. In addition, verifying that temperature-controlled perishable food are in good condition during transportation or refrigeration can be challenging to track.

A variety of smart shopping systems are presented in the literature. Wang and Yang~\cite{SysMART2:wang20163s} worked on developing a cart-based system for smart shopping. The system detects the users' presence using sensors connected to the cart's handle and collect their position, then provides promotional info for products related to their position. The system uses a mesh of wireless routers to detect position and transmit data, and a system onboard the cart to display information. The system has four types of operations including behavioral analysis, where the cart senses the actions done by the customer and provides information based on the actions. In addition, operations include Query and Answer, where the customer uses the devices LCD screen to get information. The third type is asking for help where the customer can get assistance from the supermarket's employee by requesting help from the system. The last type is equipment examination where the system probes the cart to check its status whether its malfunctioning or not being used.

Another cart-based smart shopping is developed by Alkhalawi et al.~\cite{SysMART2:alkhalawi2014smartcart}. They implemented Smart Cart, an automated personal guidance shopping system. They used RFID tags placed over the supermarket's aisles for positioning with an RFID reader attached to the cart. The authors implemented two main features: "Find your way" and "Track your trolley". The first is to locate items in the supermarket, the latter for tracking companion's cart.

Detecting the position is also achieved by Philips in a supermarket. The author in~\cite{SysMART2:lamonica2014philips}
discusses Philips' smart light system that uses the Li-Fi technology or Visible Light Communication to provide consumers with useful information based on their location inside the store by using the existing lighting fixtures. The system uses the lighting of the supermarket to transmit data to smartphones' camera at high frequency undetected by the eye thus making it easier to implement. Several other companies are developing similar systems. 

Other options for positioning includes GPS, Wi-Fi, cellular signals, Q-Track, Ultra-wideband (UWB) and RFID. Positioning systems can rely on triangulation to increase their accuracy~\cite{SysMART2:liu2007survey}. In~\cite{SysMART2:schneider2013you}, the author compares the current existing positioning technologies for indoor usage. The GPS offers a low accuracy of 10 meters but fails indoor. Wi-Fi and cellular have an accuracy of 2 meters, while Q-Track has an accuracy of 15 cm outdoor and 40 cm indoor. The UWB uses low power consumption and has an accuracy of 10 cm, but can be easily jammed indoor as well as signal bouncing problems. In addition, the authors in~\cite{SysMART2:anderson2014internet} describe existing problems with current RFID systems and survey potential solutions for proximity detection to use for positioning.

RFID are also used to track objects in real time. In~\cite{SysMART2:strickland2013tracking}, the author discusses the advantages of using RFID tags to track food from the farm to the consumer. Instead of destroying all the products due to contamination, it can be tracked to the source of contamination and decrease the amount of wasted resources. RFID tags can be designed to track the location and time of loading and unloading products during the distribution process from the processing plant to the grocery store. The problem with RFIDs is their cost and the need of compatible implementations to read the data stored by different manufacturers and users. The RFIDs can become more attractive when combined with battery-operated circuitry to log temperature and other data, and estimating the expiry date of products. Moreover, fewer people will get food poisoning as a result of their awareness of product's expiry date. 

In~\cite{SysMART2:zhou2014parking}, the authors present the implementation of Virtual Instrumentation (VI) based system used for remote monitoring of selected environmental parameters: humidity, temperature, light intensity and methane. Distance operation of the application is available via iOS apps. The authors used Dynamic Near Field Communication (DNFC) to transmit data which tracks environmental parameters of desired objects and display it in real-time. The authors in~\cite{SysMART2:yiqi2014zigbee} state that with the expansion of Internet-of-Things (IoT), smart systems are trending in the market. Remote access and control presents a challenge for smart home. To solve the challenge, the authors propose a smart home monitoring system which supports data transmission between local ZigBee network and remote Internet network. Another attempt to control devices was done by Damaj et al.~\cite{SysMART2:damaj689computer} by implementing a hardware interface connected to a computer that allows controlling external devices such as doors or televisions via a web interface accessible using mobile phones.

In this paper, we present an indoor system for connected smart supermarkets---SysMART. SysMART Indoor services aim at making shopping process simple and safe using IoT. In addition, SysMART provides indoor navigation by automatically monitoring the cart location, searching for alternative branches, checkout lane improvement based on number of items to be bought, and food safety using a DNFC tag to monitor the status of perishable goods. This paper explores the system implementation and the components used, then evaluates and analyzes it is effectiveness.

This paper is organized so that Section II presents the system design, organization and architecture. Section III presents the system implementation. A thorough analysis and evaluation with a deployment example are presented in Section IV. Section V concludes the paper and sets the ground for future work.

\section{SysMART Organization and Architecture}
\begin{figure*}[!t]
\centering
\includegraphics[width=\linewidth]{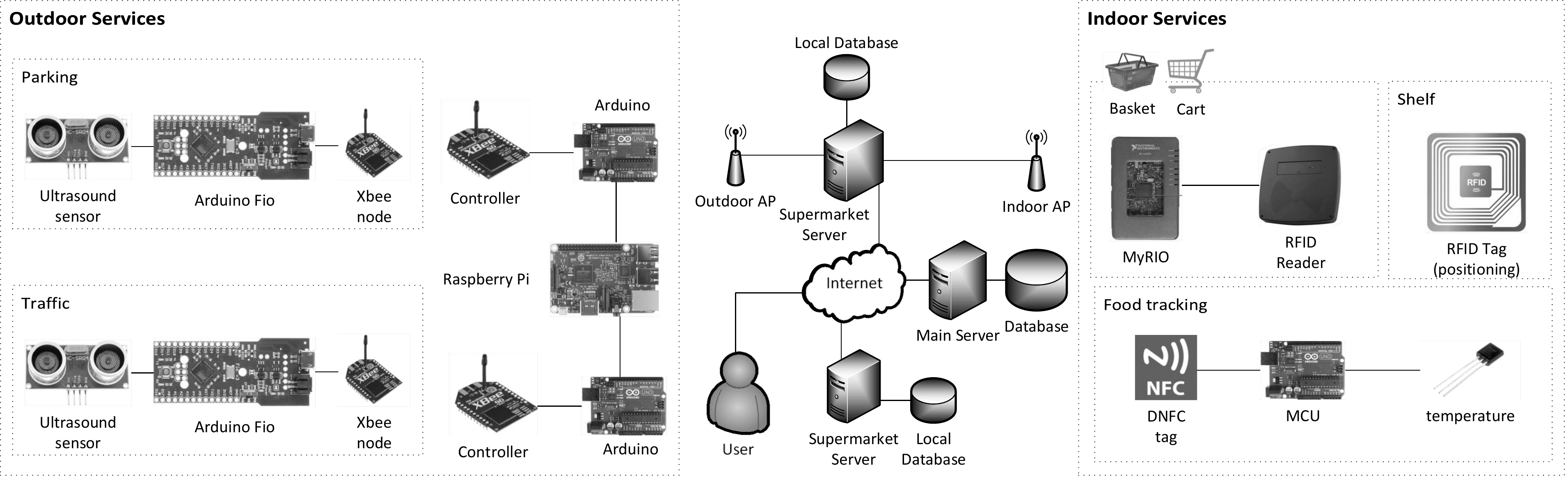}
\caption{SysMART System Architecture: To the left the Outdoor Services, in the middle the backend and to the right the Indoor Services}
\label{fig:architecture}
\end{figure*}

SysMART consists of three subsystems (see \figurename~\ref{fig:architecture}), namely, Indoor, Food Tracker and Outdoor~\cite{SysMART2:mohamad2017sysmart}. SysMART includes Local and Main servers. The Indoor subsystem is responsible for providing services inside the supermarket such as navigation, finding items and checking-out. The Food Tracker is responsible of providing status information about perishable goods. The Outdoor subsystem takes care of customers before reaching the store.

The Local server is used to store data related to a supermarket's branch and push the data to the Main server which is responsible for communicating with the customer over a mobile device. The Main server's database (\figurename~\ref{fig:database}) holds all the functional data for SysMART. The database consists of tables for inventory, products, store and other information where it gets update periodically from branches' Local server. The store table contains store location, traffic and parking status. In addition, the inventory table provides a count and availability of products. Finally, the location table stores carts position.

The Indoor subsystem provides the customer with the necessary tools to improve their shopping experience inside the supermarket. The Indoor subsystem features navigation using a pair of RF tags and RF reader mounted on the cart and sends the cart position to the Local server over Wi-Fi, fastest-check lane, alternative branches for out of stock items, requesting help on the move and reporting damaged cart using and Android application.

The Food Tracker subsystem allows customers to have a summary of perishable products status, such as, production date, expiry date, temperature and humidity info, timestamp for each distributing plant. The Food Tracker allows the customer to verify the condition in which the product was stored in. In addition, it can provide a log of all recorded data to the supermarket management or health department.
 
The Outdoor subsystem is capable of checking the parking lots status using a sensor and updates the Local server data using a mesh wireless network. A similar system is used for detecting traffic status. It checks for products' availability in a specific branch or query all branches by communicating with the main server.

\section{SysMART Implementation}
The hardware used is categorized under two subsystems: Indoor and Food Tracker. Starting with the Indoor subsystem which requires the customer to have a cart and a smartphone to get its services. For indoor navigation, a Low Frequency (LF) RFID tags rated at 125 kHz are used to identify the location. An LF RFID reader is attached to the bottom of the cart to read the tags. Although the standard LF RFID has a range of few centimeters, non-standard implementation can reach up to 1 meter under controlled conditions. Due to the cart metallic frame, the 1-meter range is reduced to 20 cm. The RFID reader is connected to a National Instruments (NI) myRIO embedded hardware device using Wiegand interface which uses 2 data lines D0 for 0-bit and D1 for 1-bit. The Ni myRIO communicates with a controlling server over Wi-Fi.

The Food Tracker subsystem makes use of Near Field Communication (NFC) available on most smartphones. The Food Tracker is a DNFC tag, unlike regular NFC tag, the data stored in the DNFC can be updated using an attached microcontroller. The updating feature makes the DNFC suitable for logging applications. The microcontroller selected is a Texas Instruments (TI) MSP430FR5969. In addition, two types of sensors are used to get the temperature and humidity readings. The temperature sensor has a high accuracy of 0.5 \textdegree{}C and low current usage rated at 7 \textmu A active mode, and 1 \textmu A sleep mode. Similarly, the humidity sensor has an accuracy of 3\% and low current usage ranging from 200 nA (sleep mode) and 820 nA (active mode). Both sensors use I2C for communication.

\begin{figure}[!t]
\centering
\includegraphics[width=\linewidth]{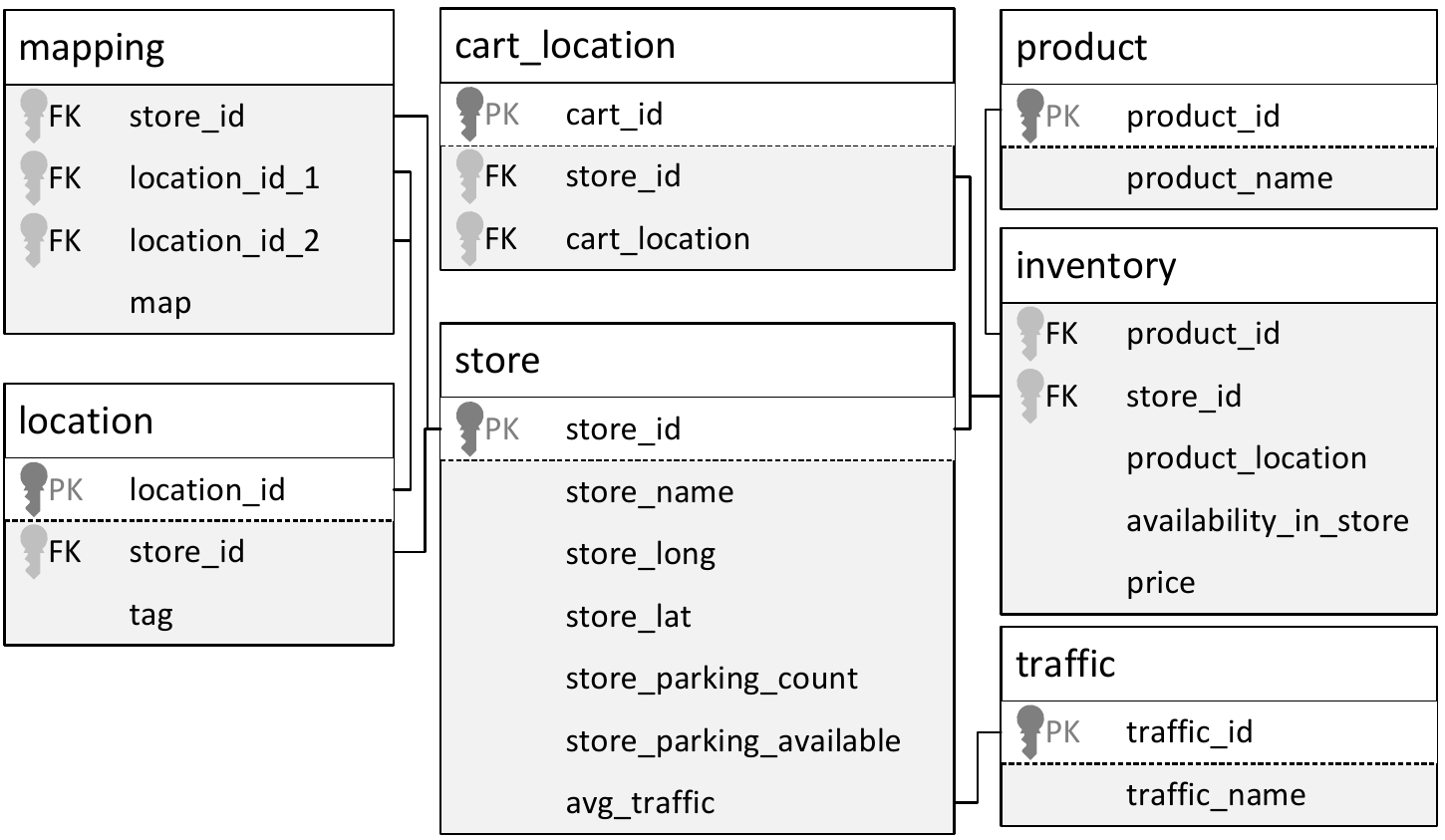}
\caption{Backend Database Schema}
\label{fig:database}
\end{figure}

Similar to the hardware subsystems, the software has several components: Backend, Indoor and Food Tracker. The backend of the project is Microsoft Azure, on which a database and a webservice are deployed to store the store data, traffic and parking status, indoor location, and other information. The information is retrieved from the database through the webservice as well. \figurename~\ref{fig:database} presents the database design. Store table provides store\_id primary and unique key for each supermarket branch.The store\_id key is used to link entries in location, inventory, cart\_location and mapping tables with each store. Each product has a product\_id  key in product table used in the inventory table to list the supermarket's inventory. The location\_id key in location table stores location id specific for each store and used to map location of inventory product using the product\_location field in inventory table and customer location for indoor navigation using the cart\_location field in cart\_location table.

The second component is the indoor software subsystem, which is made by developing a myRIO program and an Android application. The myRIO program is developed using NI LabVIEW a visual programming language. The deployed software reads the location tag ID from the RFID reader using Wiegand 26 protocol. The application server can only receive data from one myRIO at a time, to avoid having two carts sending data at the same time, a random delay is introduced before transmitting to minimize data collision. The data transmitted consists of 16-bit integer store ID, 16-bit integer cart ID, and a 6 hexadecimal characters for tag ID, totaling to 80 bits with overhead may reach up to 128 bits. The used Wi-Fi transmits at 54 Mbps, therefore, each data sent will take 2.26 \textmu s or 442,368 data/s, making the random delay up to 1 second adequate to avoid data collision.

\begin{figure*}[!t]
\centering
\includegraphics[width=0.7\linewidth]{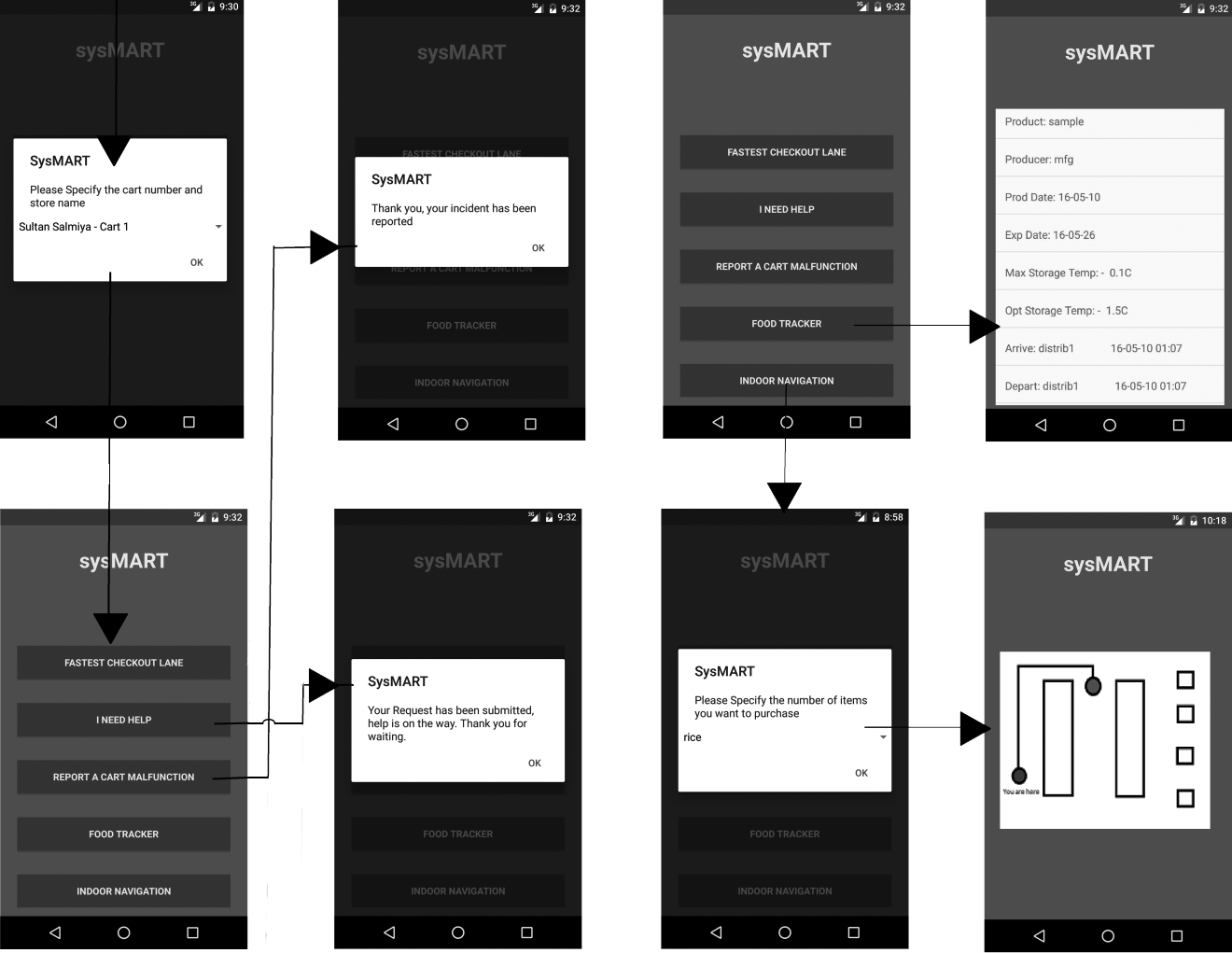}
\caption{SysMART Indoor App Activities and Transactions}
\label{fig:app-flow}
\end{figure*}

The application server is developed using LabVIEW and can run on a personal computer or a tablet. The app server's main functionality is to update the database using the data it receives from the myRIOs deployed on the carts. A second functionality is to display the request of assistance and malfunctioning carts. The user interaction with the cart is accomplished by using an Android app developed using Android Studio. The customer enters the store ID and cart ID to get the related data of the cart. The app provides the current location and the location of the item desired (see \figurename~\ref{fig:app-flow}). Using the app, customers can get the fastest lane for checkout. The app also provides the ability to request assistance and report cart malfunctioning.

The last component is the Food Tracker. The Food Tracker tag is developed using TI Code Composer Studio (CCS) which allows to build and debug the code. As the data stored in the tag are sensitive by nature, such as, production data and expiry date. The data must be tamper proof, therefore the code must not allow modifying any pre-populated field without resetting the tag. Moreover, resetting the tag introduces multiple usages of the tag. An additional security measure is a 20-character password provided to the tag at initialization time after a reset. The reset can be performed after confirming the password and it clears the settings. To avoid tampering with time logging, all timestamps are provided by the microcontroller's real-time-clock at the time of setting a field (when a product arrives at a distributing plant or departs from one). For temperature and humidity logging, due to limited space in the microcontroller's memory, the data logged after the difference between a new reading and the last recorded reading exceeds a specific threshold. Also to minimize battery consumption and writing space, the sensor readings are set at specific intervals. The communication with the tag is done over NFC. The DNFC reads or writes data, and after the reader/writer stops the signal, the DNFC notifies the microcontroller of an action was done. 

\begin{figure}[!t]
\centering
\subfloat[Read Tag]{\includegraphics[width=0.48\linewidth]{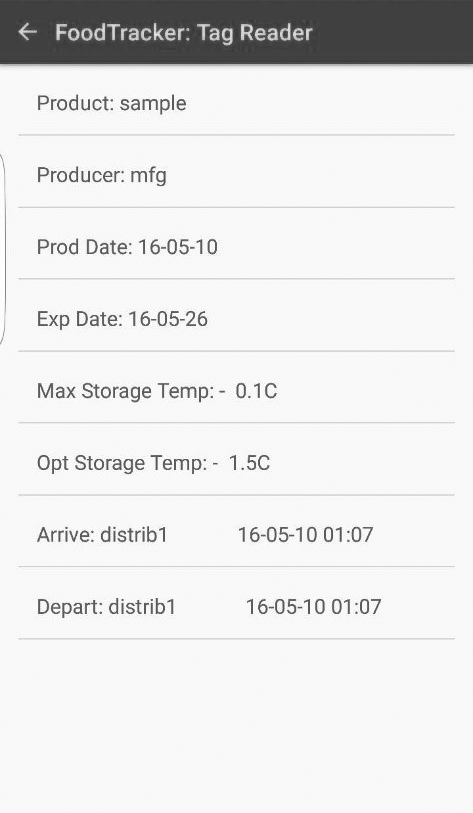}
\label{fig:admin-app-read}}
\hfill
\subfloat[Graph Data]{\includegraphics[width=0.48\linewidth]{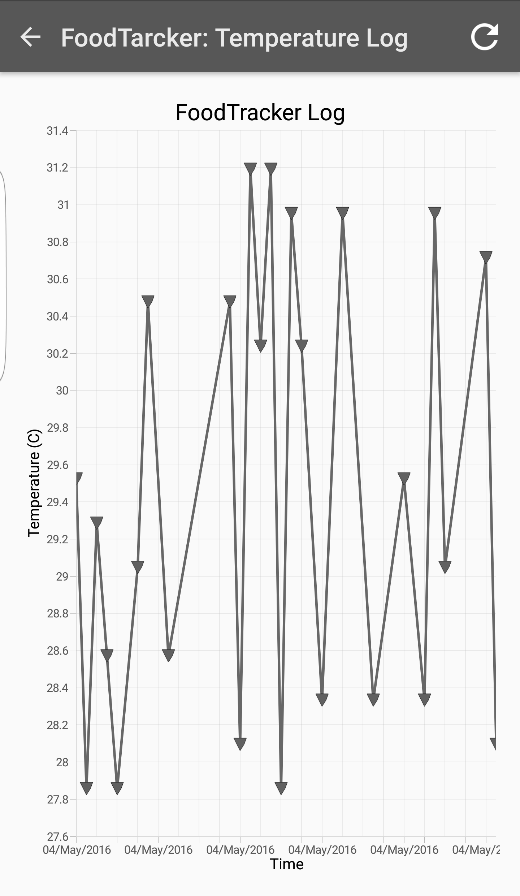}
\label{fig:admin-app-graph}}
\caption{Food Tracker Master App}
\label{fig:admin-app}
\end{figure}

By default, the DNFC holds the latest summary data in its memory. The microcontroller is Byte addressable; to avoid data corruption each log item should be stored separately. The log item contains the timestamp, temperature and humidity value in raw binary format, this will reduce the space needed to store the data. The timestamp for the log item is an increment of the first log item's value, so the first item will have a timestamp of 0. The temperature sensor value occupies 12 bits and the humidity sensor occupies 14 bits for the highest possible accuracy, the total sensors bit count is 26 bits. The next 8-bit multiples are 32, 40 and 48. Choosing the 48-bit format provides 22 bits for timestamp, with increment of 1 minute. Thus the timestamp allows for a maximum of 7.98 years of log tracking. To communicate with the tag, a master application is developed for Android phones. The application provides the capabilities of setting the tabs fields, reading the summary info and plot a graph from the detailed log as shown in \figurename~\ref{fig:admin-app}.

\section{Analysis and Evaluation}
SysMART is built and realized successfully and proves to be effective in application. First, the goal of SysMART is to help the society and economy. By connecting the stores, the customer can have access to information from other stores.  Customers should have an account with the supermarket to facilitate the interaction with the store, to locate friends and relatives and other features. By having an account, a payment through the account may be possible at in the future. The aforementioned features raise a concern regarding privacy, security, and safety. Customers’ information are very sensitive and must be protected including their location and they must have an option to disable storing their location. Since shopping experience becomes improved as a result of the project, customers may become more inclined to spend more, therefore, affecting the purchasing power.

\begin{figure}[!t]
\centering
\includegraphics[width=\linewidth]{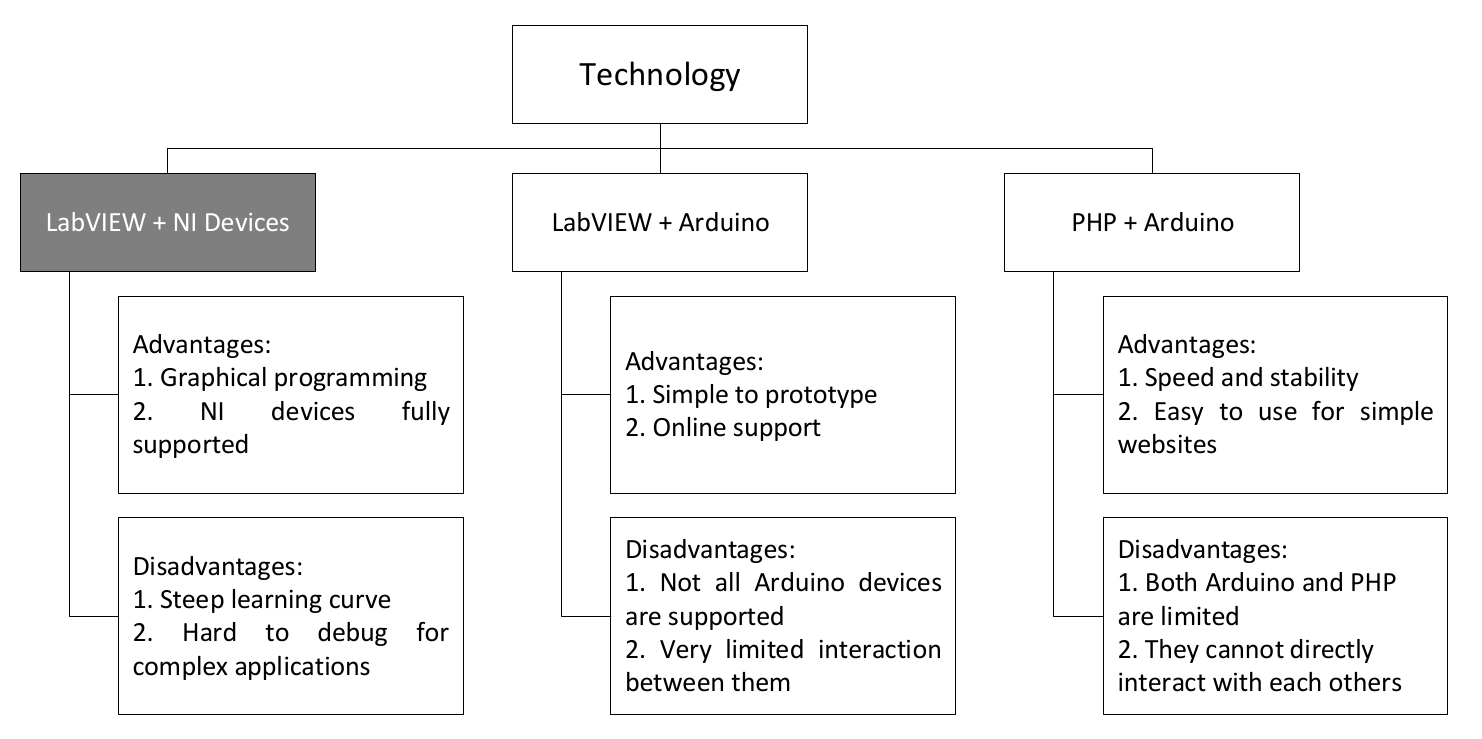}
\caption{Technologies}
\label{fig:tree-technology}
\end{figure}

Another feature, which raises concerns, is the food tracking. The tracking tags must be reusable, to reduce additional waste and extra resources. The addition of a tracker may also increase the price tag of the product from the tracker cost and extra processing overhead that comes with it. The components used were chosen based on availability, performance, scalability and cost. From the tree (\figurename~\ref{fig:tree-technology}) the most promising technology for our purposes is the LabVIEW and NI devices. Although the decision is to use NI devices, but other microcontrollers for specific purposes.

The RFID devices are grouped by frequency. There are three main categories: Low Frequency (LF), High Frequency (HF) and Ultra High Frequency (UHF). The HF or NFC, operates at 13.56 MHz, is used in different fields, from identification, smart wallet and transmitting data and has a maximum range of 1 m. For the Food Tracker, the NFC is the most suitable solution as it allows data communication although UHF RFID provides similar functionality, the reader and antenna are very expensive, also detecting the tags from a distance may cause some confusion as to which product is being tracked. Furthermore, most Android smartphones includes and NFC readers which makes the NFC more favorable over UHF RFID.

\begin{figure}[!t]
\centering
\includegraphics[width=\linewidth]{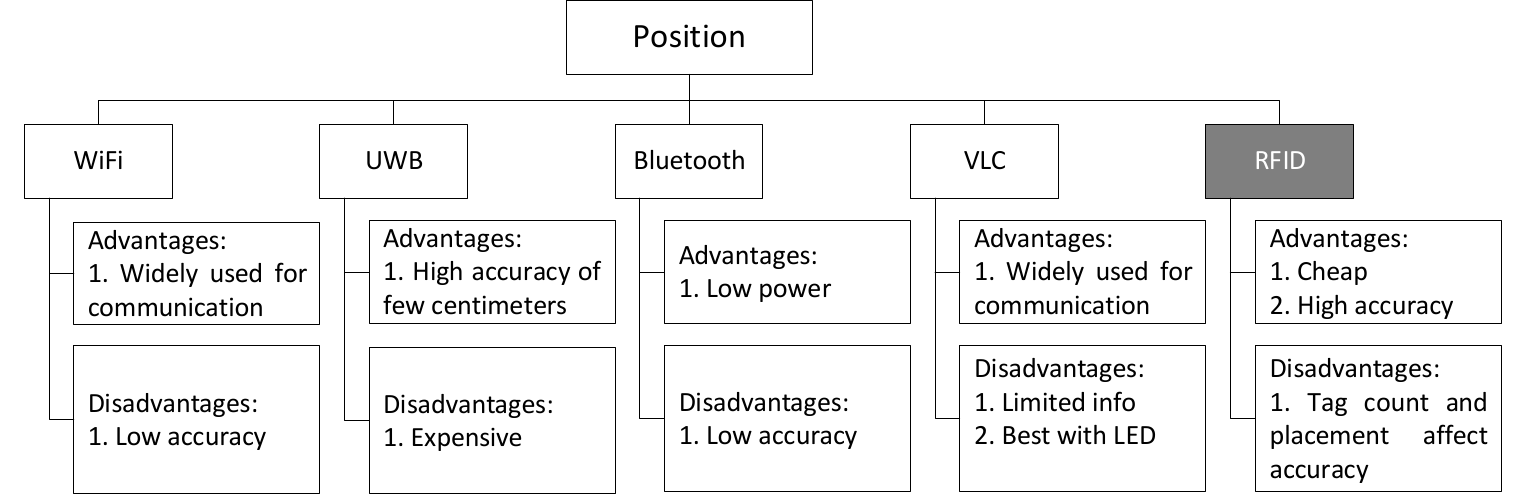}
\caption{Positioning Technology}
\label{fig:tree-position}
\end{figure}

\begin{figure}[!t]
\centering
\includegraphics[width=\linewidth]{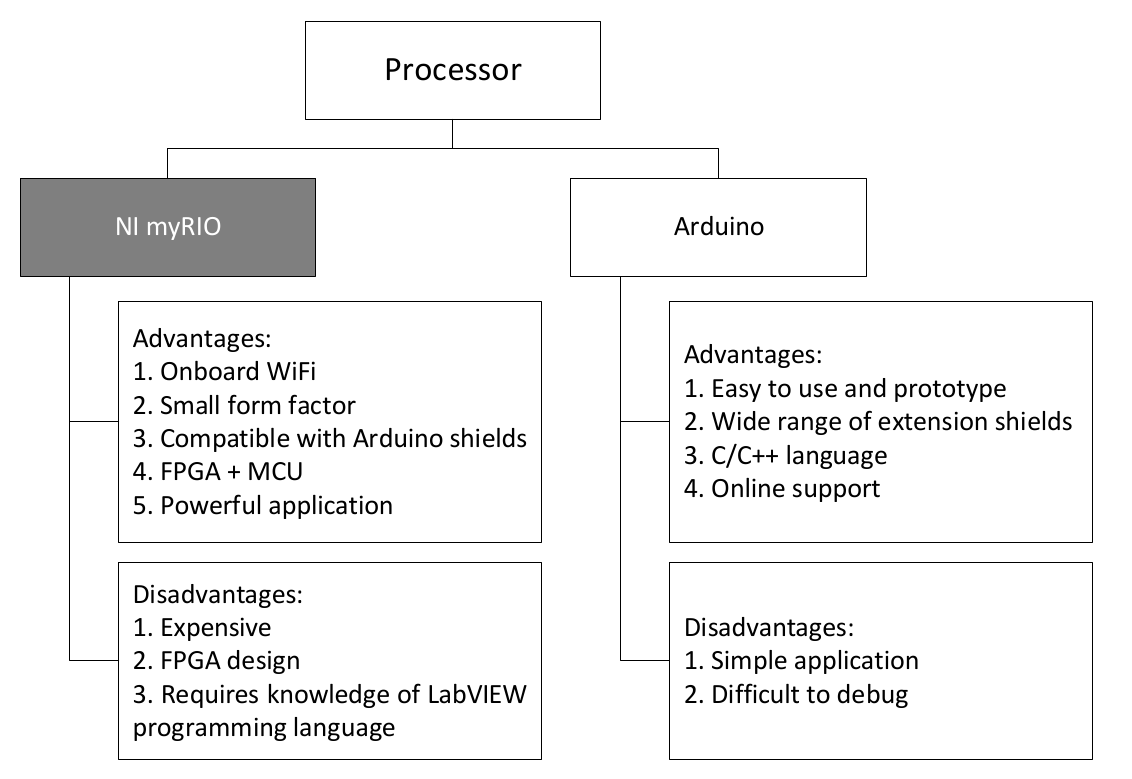}
\caption{Cart Processor}
\label{fig:tree-processor}
\end{figure}

From \figurename~\ref{fig:tree-position} both VLC and RFID are valid options; but as VLC is still a new technology, RFID tags are more favorable to implement positioning. The 10 cm NFC's range is short and requires the device to be close to the ground, and the UHF RFID  reader is more expensive than its alternatives. While the LF RFID has a short range of 10 cm, non-standard implementation can support up 80 cm using a 30 cm by 30 cm reader.

Next, deciding on processing unit to be used from \figurename~\ref{fig:tree-technology}, the NI devices for critical applications is chosen as the processing unit for the cart or basket. NI products are compared to other MCUs, such as, Arduino as shown in \figurename~\ref{fig:tree-processor}.

Food tracker is a device that allows the consumer, vendor and distributor to check the safety of the product. As the lifespan of a product can be affected by several factors, mainly temperature, so the food tracker must log the temperature regularly and inform the user of the maximum temperature reached, average temperature and estimated expiry date based on the logged temperatures. Consumers can use their smartphones to get a summary of the food status and the vendors can get a detailed log of the storage conditions and product hand-off along the distribution chain.

\begin{figure}[!t]
\centering
\includegraphics[width=\linewidth]{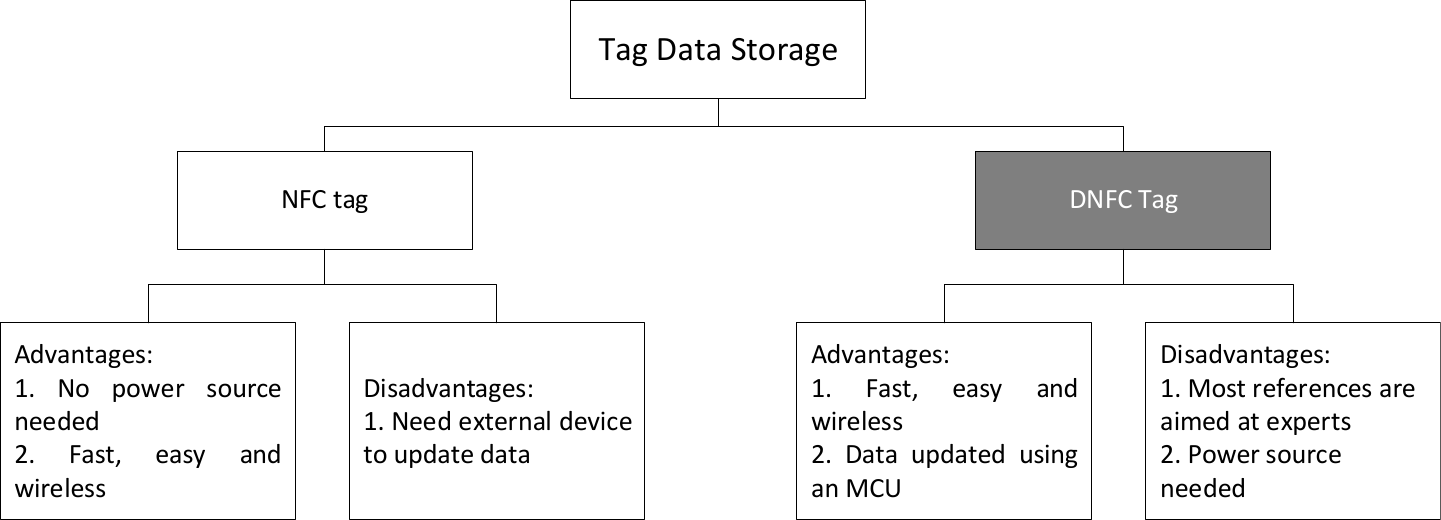}
\caption{Food Tracker Tag Type}
\label{fig:tree-data}
\end{figure}

\begin{figure}[!t]
\centering
\includegraphics[width=\linewidth]{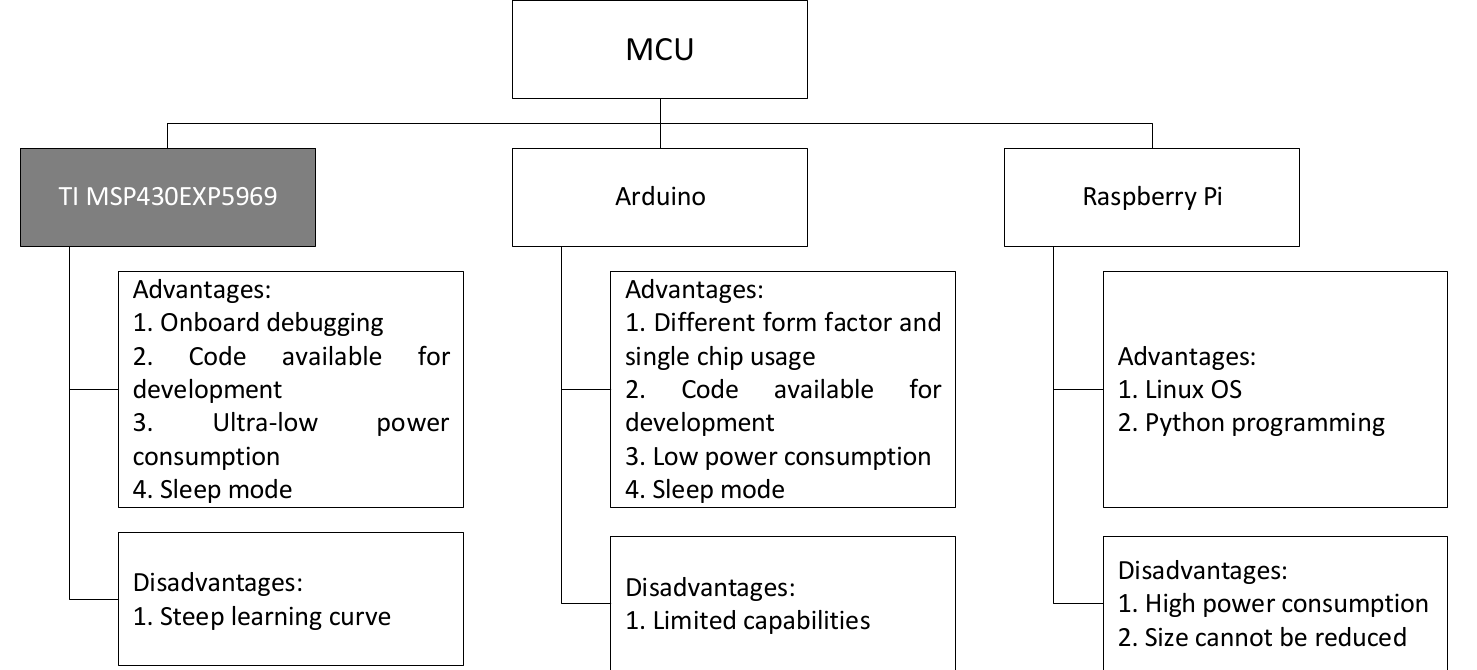}
\caption{Food Tracker Processor}
\label{fig:tree-mcu}
\end{figure}

A useful technology used to access data in a simple and fast way is NFC tags. NFC tags can be accessed wirelessly to retrieve data and are passive devices so they require no power source to be active as they become active by the power provided wirelessly by the reader device. The NFC tags can be used to store the condition of the product. But they need to be accompanied by a writer device at a regular interval to update the value, which makes them inappropriate for the application because the tag must be updated as needed and the system should be independent from external components. An alternative to the NFC tag is DNFC tag (\figurename~\ref{fig:tree-data}). The DNFC has the same features as a regular NFC tag, but with the ability to interface with a microcontroller to update the data. Since the DNFC tag needs a microcontroller, the TI microcontroller was chosen. The TI microcontroller offers on-board real time clock and an FRAM to store data. Two other alternatives were Arduino and Raspberry Pi (\figurename~\ref{fig:tree-mcu}).

\begin{figure}[!t]
\centering
\includegraphics[width=\linewidth]{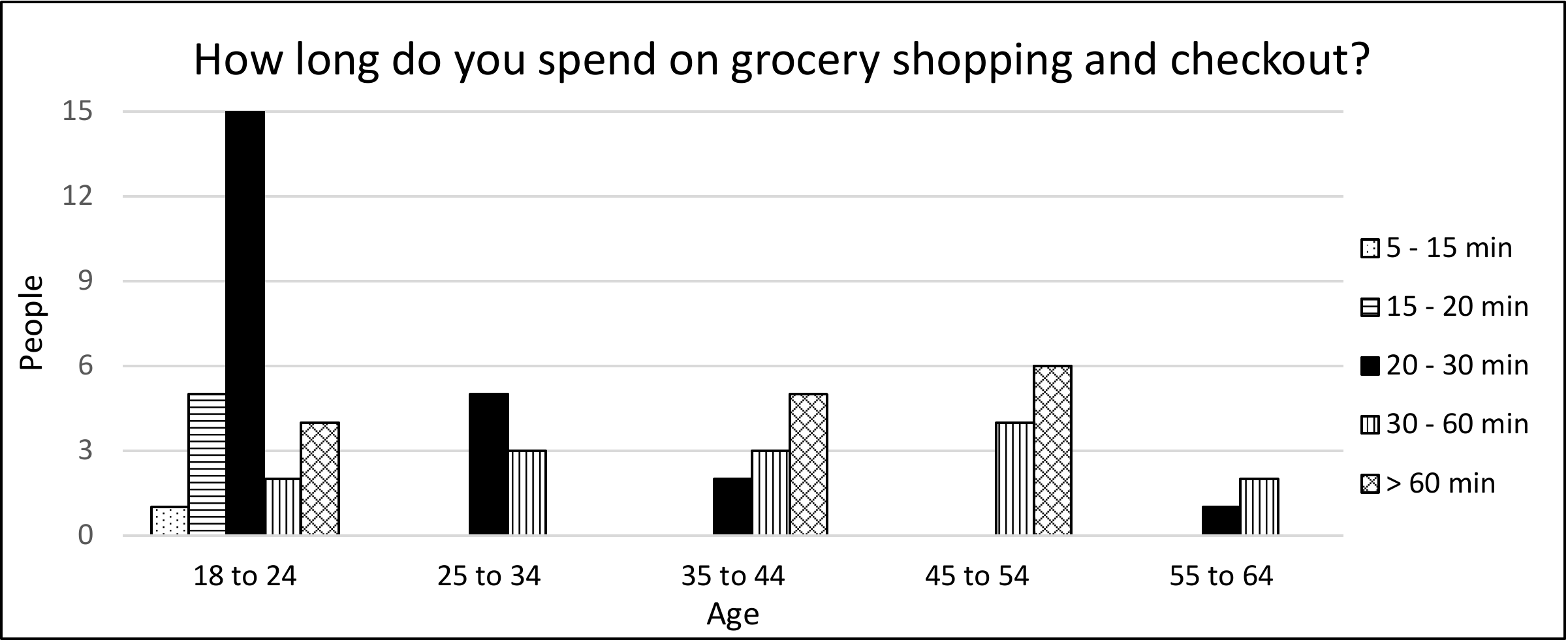}
\caption{Survey: Grocery Time}
\label{fig:survey-time}
\end{figure}

\begin{figure}[!t]
\centering
\includegraphics[width=\linewidth]{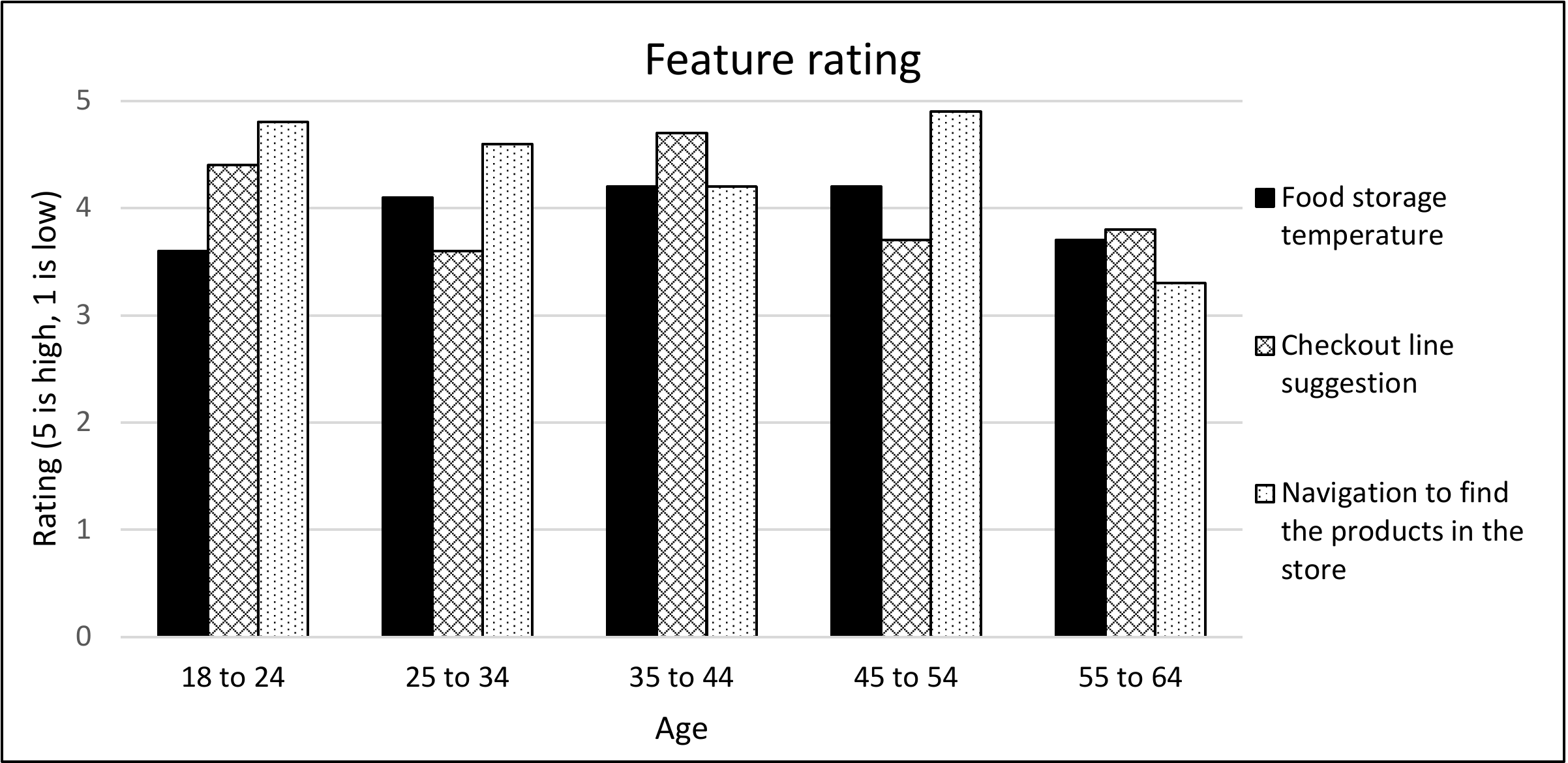}
\caption{Survey: Feature Rating}
\label{fig:survey-features}
\end{figure}

To evaluate SysMART from the user's perspective, a survey is deployed on a population of 58 people between the ages of 18 to 64. The survey collects the average time of shopping and the feature ratings from customers' point-of-view. \figurename~\ref{fig:survey-time} and \figurename~\ref{fig:survey-features} provide a bar-graph of the survey's results.

To estimate the cost and feasibility of SysMART, a simulated case study was applied to a local supermarket chain. The supermarket chain owns several branches and the simulation was applied to one of the busiest mostly frequented branch. The supermarket has 150 carts. A total of 230 RFID tags are carefully distributed among different sections. An overall budget for the simulated case study is \$88,730 and \$15 per food tag.

\section{Conclusion}
SysMART is a modern IoT system that offers fast and safe shopping experiences. SysMART supports a bouquet of features that include indoor navigation, fast checkouts, and food tracking. SysMART interacts with the customers' smartphone to provide real-time information. The cost associated with offering a premium service to customers is expected to have a high return on investment--with more customers visiting the supermarket for efficient grocery shopping and checkout. Future works include motorizing and tracking the cart to allow smoother shopping for elderly and kids, and facilitate transportation of heavy items. Moreoever, future work includes accelerating security aspects and database queries using high-performance computing~\cite{SysMART2:damaj2007parallel,SysMART2:damaj2001performance,SysMART2:damaj2001performance2, SysMART2:singh2014survey}.


\bibliographystyle{IEEEtran}
\bibliography{IEEEabrv,SysMART2}

\end{document}